\documentstyle[12pt]{article}
\begin{document}
\def\beq{\begin{equation}}
\def\eeq{\end{equation}}
\def\bea{\begin{eqnarray}}
\def\eea{\end{eqnarray}}
\def\ve{\vert}
\def\vel{\left|}
\def\ver{\right|}
\def\nnb{\nonumber}
\def\ga{\left(}
\def\dr{\right)}
\def\aga{\left\{}
\def\adr{\right\}}
\def\rar{\rightarrow}
\def\nnb{\nonumber}
\def\la{\langle}
\def\ra{\rangle}
\def\ba{\begin{array}}
\def\ea{\end{array}}
\def\tep{$B \rar K \ell^+ \ell^-$}
\def\tepm{$B \rar K \mu^+ \mu^-$}
\def\tept{$B \rar K \tau^+ \tau^-$}
\def\ds{\displaystyle}

\title{
{\bf \Large Radiative $B_c \rar \tau \bar \nu_\tau \gamma$ decay } }

\author{\vspace{1cm}\\
{\small T. M. Aliev \thanks
{electronic adress: taliev@rorqual.cc.metu.edu.tr}\,\,,
M. Savc{\i} }\thanks
{electronic adress: savci@rorqual.cc.metu.edu.tr} \\
{\small Physics Department, Middle East Technical University}, \\
{\small 06531 Ankara, Turkey } }

\date{}

\begin{titlepage}
\maketitle
\thispagestyle{empty}

\begin{abstract}
\baselineskip  0.7cm
The radiative $B_c \rar \tau \bar \nu_\tau \gamma$ decay is analyzed in the 
standard model. The branching ratio of this decay is calculated and 
the contributions of the Bremstrahlung and structure dependent parts 
are compared. It is shown that this decay can be detected at LHC.

\end{abstract}

\vspace{1cm}
~~PACS number(s): 13.25.Gv, 13.20.--v, 13.10.+q

\end{titlepage}

\section{INTRODUCTION}
Recently the CDF collaboration at Fermilab Tevatron in $1.8~\mbox{\rm TeV} ~p \bar p$
has reported the discovery of $B_c$ meson
with ground state mass $m_{B_c} = 6400 \pm 390 \pm 130$ MeV and life time
$\tau(B_c) = 0.46~ ^{+0.18}_{-0.16}~ps $ \cite{R1}. This discovery has stimulated 
the investigation of the properties of the $B_c$ meson theoretically, as well
as experimentally, on a new footing. $B_c$ meson contains two heavy quarks 
and for this reason the perturbative QCD predictions are more reliable.
Therefore the study of the decays of the $B_c$ meson allows to check 
the QCD predictions more precisely and one can get 
essential new information about the confinement scale inside hadrons.  

The weak decay
channels of the $B_c$ meson are richer than the corresponding $B_q\,(q =
u,~d,~s)$. The weak $B_c$ meson decay channels can
be divided into three classes: a) The $\bar b$ quark decay with the
spectator $c$ quark, for example , $B_c \rar J/\psi \ell \bar \nu_\ell$;
b) the $c$ quark, decay with the spectator $\bar b$ quark, for example,
$B_c \rar B_s \ell \bar \nu_\ell$ and c) the annihilation channels like  
$B_c \rar \ell \bar \nu_\ell\,(c\bar s,\,u \bar s)$, where 
$\ell = e,\,\mu,\,\tau$. 

From experimental point of view, investigation of the weak decays of the 
$B_c$ meson
gives us the most direct information in determining corresponding elements
of the Cabibbo--Kobayashi--Maskawa matrix (CKM), such as $V_{cb}$ etc.
A comprehensive analysis of the $B_c$ meson spectroscopy and strong and
electromagnetic decays of the excited states is given in \cite{R2}.
The semileptonic and various nonleptonic decays and the $B_c$ life time have
been calculated in many works (see for example \cite{R3}--\cite{R10} and
references listed therein).

The pure
leptonic decays of $B_c$ meson are the simplest among all decays and these
decays can be useful in determination of the leptonic decay constant
$f_{B_c}$ of the $B_c$ meson. The 
$B_c \rar \ell \bar \nu_\ell~(\ell = e,\,\mu)$
decay is helicity suppressed and it makes the determination of 
$f_{B_c}$ of the $B_c$ meson very difficult.
Although the $B_c \rar \tau \bar \nu_\tau$ channel is free of the helicity
suppression, its observation is possible if we have a good efficiency for
detection of the $\tau$ lepton. In this work we investigate the 
$B_c \rar \tau \bar \nu_\tau \gamma$ decay in frame work of the standard
model. In $B_c \rar \tau \bar \nu_\tau \gamma$ decay
comparable contributions come from diagrams where photon is emitted both
from initial quark lines and final $\tau$ leptons. These contributions can
give very useful information about relative roles of the strong and
electroweak interactions. Note that in the
$B_c \rar \ell \bar \nu_\ell \gamma~(\ell = e,\,\mu)$ decay, it follows from
helicity arguments that the contribution of the diagram where photon
radiated from charged lepton, must be proportional to the lepton mass
$m_\ell$, and hence it can safely be neglected (see \cite{R11, R12}).

In \cite{R13,R14} the number of $B_c$ mesons that will be produced in LHC is
estimated to be $\sim 2 \times 10^8$. This clearly is an indication of the
real possibility of an experimental investigation of the
properties of the $B_c$ meson at LHC.
This paper is organized as follows. In section 2 we give necessary
theoretical formalism for the $B_c \rar \tau \bar \nu_\tau \gamma$ decay.
Section 3 is devoted to the numerical analysis and the discussion of the
results.  

\section{FORMALISM FOR THE $B_c \rar \tau \bar \nu_\tau \gamma$ DECAY}
The matrix element for the $B_c \rar \tau \bar \nu_\tau \gamma$ decay can be
obtained from $B_c \rar \tau \bar \nu_\tau$ process if the photon is
attached to any charged fermion line. The effective Hamiltonian for the 
$B_c \rar \tau \bar \nu_\tau$ decay is
\bea
{\cal H} = \frac{G}{\sqrt{2}}\,V_{cb}\, \bar c \gamma_\mu
\ga 1 - \gamma_5 \dr b \,\bar \tau \gamma_\mu \ga 1 - \gamma_5 \dr
\nu_\tau~.
\eea

Let us start our investigation for the process 
$B_c \rar \tau \bar \nu_\tau \gamma$ by considering the Bremstrahlung 
diagram, where photon is radiated from $\tau$ lepton. The
corresponding matrix element is 
\bea
{\cal M}^{Brem.} = i\, f_{B_c} e \,\frac{G}{\sqrt{2}} \,V_{cb} \varepsilon_\alpha
\bar u(p_1) \left\{ \gamma_\alpha + \frac{m_\tau}{2 p_1 q}
\left[ 2 p_{1\alpha} + \gamma_\alpha \not\!q \right] \right\}
\ga 1 - \gamma_5 \dr v(p_2)~,
\eea
where $\varepsilon_\alpha$ and $q_\alpha$ are the photon four--vector
polarization and momentum, $p_1$ and $p_2$ are the momenta of
$\tau$ lepton and neutrino, respectively, and $f_{B_c}$ is the leptonic
decay constant of $B_c$ meson, and defined as
\bea
\left<0\vel \bar c \gamma_\mu \gamma_5 b  \ver B_c \right> = - i
f_{B_c} P_{B\mu}~ \nnb.
\eea
From Eq. (2) we immediately see that the matrix element ${\cal M}^{Brem.}$
is not gauge invariant. 
In regard to the contribution of the structure dependent part to the $B_c
\rar \tau \bar \nu_\tau \gamma$ process, i.e., when photon is radiated 
from initial quark lines, we consider the following correlator:
\bea
{\cal M}_\mu^{SD} = \frac{G}{\sqrt{2}} \,V_{cb}\varepsilon_\alpha
\int d^4 x e^{i q x} \left<0 \vel {\cal T} \left\{ \bar c(0) \gamma_\mu
\ga 1 - \gamma_5 \dr b(0) J_\alpha^{el} (x) \right\} \ver B_c \right>~,
\eea
where $J_\alpha^{el}$ is the electromagnetic current for $b$ or $c$ quarks.

Vector part $\ga {\cal M}^V \dr$ of this correlator is obviously equal to 
\bea
{\cal M}_\mu^V = e \,\frac{G}{\sqrt{2}} \,V_{cb} \, \frac{f_1(p^2)}{m_{B_c}^2}
\epsilon_{\mu\alpha\rho\beta} \varepsilon_\alpha p_\rho q_\beta~,
\eea
where $f_1(p^2)$ is the transition form factor, and $p$ is the momentum
transfer, $p = p_1 + p_2 = p_1 - q$. 
In general axial part of the correlator can be written as
\bea
{\cal M}_\mu^A = \frac{G}{\sqrt{2}} \,V_{cb}\varepsilon_\alpha
\left\{ A_1 \, g_{\mu\alpha} + A_2 \, p_\mu p_\alpha + A_3 \, p_\alpha q_\mu
+ A_4 \, p_\mu q_\alpha + A_5 \, q_\mu q_\alpha \right\}~,
\eea
where $A_i$ are the form factors. Since $q \varepsilon=0$, the form factors
$A_4$ and $A_5$ can be omitted. Obviously, to obtain the matrix element
which describes the structure dependent part ${\cal M}^{SD}$ we must
multiply Eqs. (4) and (5) to the leptonic current. From Eqs. (5) it
follows that ${\cal M}^A$ part is not gauge invariant as well, 
while ${\cal M}^V$ is gauge invariant itself. Since the total matrix element
is supposed to be gauge invariant, we demand that 
${\cal M}^{Brem.} + {\cal M}^A$ must be gauge invariant, i.e.,    
\bea
q_\alpha\left\{ {\cal M}^{Brem.}_\alpha + {\cal M}^A_\alpha \right\}= 0 ~.
\eea
This condition allows us to find the relation between different form factors.
Indeed from Eqs. (2) and (5) (after multiplying Eq. (5) to the leptonic
current), we have
\bea
i f_{B_c} e \, \bar u \ga \not\! q + m_\tau \dr
\ga 1 - \gamma_5 \dr v + \bar u \left[ A_1 + A_2 \, (p q) \right] \not\! q
\ga 1 - \gamma_5 \dr v +
A_3 \,(p q) m_\tau \bar u \ga 1 - \gamma_5 \dr v = 0~.
\eea
From this equation we have
\bea
A_1 + A_2\,(p q) &=& - i f_{B_c} e~, \\
A_3\,(p q)  &=& - i f_{B_c} e~.
\eea
If we set $p q = 0$ in Eq. (8), we get 
$A_1(p q = 0 ) = - i f_{B_c} e$. So the general form of $A_1$ is as follows:
\bea
A_1 = - i f_{B_c} e + b (p q) ~,
\eea
where $b$ is a new, unknown form factor. It follows form Eqs. (8) and (10)
that
\bea 
A_2 = -b~. \nnb
\eea
In the same manner, from Eq. (9) we get 
\bea 
A_3 = - \frac{i f_{B_c} e}{p q}~. \nnb
\eea
Substituting these expressions for $A_1.~A_2$ and $A_3$ in the Bremstrahlung
and axial parts of the matrix element, we get
\bea 
{\cal M}^{Brem.} + {\cal M}^A &=& \frac{G}{\sqrt{2}} \,V_{cb} e \Bigg\{
i f_{B_c} m_\tau \bar u(p_1) \left[ \frac{p_1 \varepsilon}{p_1 q} - 
\frac{p \varepsilon}{p q} \right] \ga 1-\gamma_5  \dr v(p_2) \nnb \\
&+& \!\! i f_{B_c} \frac{m_\tau}{2 p_1 q} \bar u(p_1) 
\not\!\varepsilon \not\! q \ga 1-\gamma_5  \dr v(p_2) \nnb \\
&+&\!\! i\, \frac{f_2(p^2)}{m_{B_c}^2} \left[ \varepsilon_(p q) -
q_\mu (p \varepsilon) \right] \bar u(p_1) \gamma_\mu 
\ga 1-\gamma_5  \dr v(p_2) \Bigg\}~,
\eea
where for convenience we have redefined $b(p^2)$ as  
\bea 
b(p^2) = i \, \frac{f_2(p^2)}{m_{B_c}^2}. \nnb
\eea
The second term in the first line of the Eq. (11), corresponds physically to
the contribution of the one $B_c$ intermediate state.
It easy to check that the gauge invariance of the sum of the Bremstrahlung and
axial parts of the matrix element in Eq. (11) is reestablished. Adding the
vector part of the correlator to Eq. (11), we finally get for the matrix
element of the $B_c \rar \tau \bar \nu_\tau \gamma$ process
\bea
{\cal M} = {\cal M}_1 + {\cal M}_2 ~, \nnb
\eea
where
\bea
{\cal M}_1 &=& \frac{G}{\sqrt{2}} \,V_{cb} e \Bigg\{
\frac{f_1(p^2)}{m_{B_c}^2} \epsilon_{\mu\alpha\rho\beta} 
\varepsilon_\alpha p_\rho q_\beta  \nnb \\
&+&\!\! i \frac{f_2(p^2)}{m_{B_c}^2} 
\left[ \varepsilon_\mu (p q) - q_\mu (p \varepsilon) \right] \Bigg\}
\bar u(p_1) \gamma_\mu \ga 1 -\gamma_5 \dr v(p_2) ~,\\ \nnb \\ \nnb \\  
{\cal M}_2 &=& \frac{G}{\sqrt{2}} \,V_{cb} i f_{B_c} m_\tau \,\bar u(p_1)\Bigg\{
\frac{1}{2 p_1 q} \not\!\varepsilon \! \not\! q 
+ \left[ \frac{p_1 \varepsilon}{p_1 q} - 
\frac{p \varepsilon}{p q} \right] \Bigg\}
\ga 1 - \gamma_5 \dr v(p_2) ~.
\eea

As we noted earlier, ${\cal M}_2$ is indeed proportional to the lepton mass, 
as expected from helicity arguments.
If we formally set $m_\tau \rar 0$ in Eqs. (12) and (13), the resulting
expression is expected to coincide with the $B_c \rar \ell \bar \nu
\gamma~(\ell = e,~\mu)$. 
This decay was investigated in the framework of light cone QCD sum rules and
the constituent quark model approach in \cite{R11} and \cite{R15},
respectively.

After lengthy, but straightforward calculation for the squared matrix
element, we get
\bea
\vel {\cal M} \ver ^2 = \vel {\cal M}_1 \ver ^2 + 
2 \,\mbox{\rm Re}\left[ {\cal M}_1 {\cal M}_2^\dagger \right] 
+ \vel {\cal M}_2 \ver ^2~,
\eea
where
\bea
\vel {\cal M}_1 \ver ^2 &=& \frac{G^2}{2} \vel V_{cb} \ver^2 e^2 16
\left[ \frac{\vel f_1(p^2) \ver^2}{m_{B_c}^4} + 
\frac{\vel f_2(p^2) \ver^2}{m_{B_c}^4} \right] \nnb \\
&\times& \!\! \left\{ (p p_2) (p q)  (p_1 q) +
(p_2 q) \left[ (p p_1)  (p q) - p^2 (p_1 q) \right] 
\right\}~,\\ \nnb \\ \nnb \\
2 \,\mbox{\rm Re}\left[ {\cal M}_1 {\cal M}_2^\dagger \right] &=&
\frac{G^2}{2} \vel V_{cb} \ver^2 e^2 \ga - 16 f_{B_c} m_\tau^2 \dr 
\frac{1}{(p_1 q) (p q)}\nnb \\
&\times& \!\! \Bigg\{ \frac{ f_2(p^2)}{m_{B_c}^2} \, p^2 (p_1 q) (p_2 q) 
+ (p q)^2 \left[ \frac{ f_2(p^2)}{m_{B_c}^2} \,(p_1 p_2 + p_2 q) -
\frac{ f_1(p^2)}{m_{B_c}^2} \, (p_2 q) \right] \nnb \\
&-& \!\! \frac{ f_2(p^2)}{m_{B_c}^2} 
\left[ (p p_2) (p_1 q) + (p p_1) (p_2 q) \right] \Bigg\}~, 
\\ \nnb \\ \nnb \\
\vel {\cal M}_2 \ver ^2 &=& \frac{G^2}{2} \vel V_{cb} \ver^2 e^2 
\ga - 4 f_{B_c}^2 m_\tau^2 \dr 
\frac{1}{(p_1 q)^2 (p q)^2} \nnb \\
&\times& \!\!\Bigg\{ 2 p^2 (p_1 p_2) (p_1 q)^2 + (p q)^2 \left[
(p_1 p_2) \ga 2 m_\tau^2 - p_1 q \dr +
(p_2 q) \ga m_\tau^2 - 2 p_1 q \dr \right] \nnb \\
&+& \!\! (p q) (p_1 q) \left[ (p p_2) (p_1 q) - (p p_1)
(4 p_1 p_2 + p_2 q) \right] \Bigg\}~.
\eea
All calculations have been performed in the rest frame of the $B_c$ meson.
The dot products of the four--vectors are defined if the photon and neutrino
(or electron) energies are specified. The Dalitz boundary for the photon energy
$E_\gamma$ and neutrino energy $E_2$ is as follows:
\bea
\frac{m_{B_c}^2 - 2 m_{B_c} E_\gamma - m_\tau^2}
{2 m_{B_c}} \le \!\!\! &E_2& \!\!\! \le
\frac{m_{B_c}^2 - 2 m_{B_c} E_\gamma - m_\tau^2} 
{2 \ga m_{B_c} - 2 E_\gamma \dr}~, \nnb \\ \nnb \\
0 \le \!\!\! &E_\gamma& \!\!\! \le 
\frac{m_{B_c}^2 - m_\tau^2}{2 m_{B_c}}~.
\eea
The expression for the differential decay rate can be written as
\bea
\frac{d\Gamma}{d E_2\,d E_\gamma} = \frac{1}{64 \pi^3 m_{B_c}}
\vel {\cal M} \ver^2~.
\eea

The differential $\ga d\Gamma/d E_\gamma \dr$ and total decay width 
are singular at the lower limit of the photon energy, and this singularity
which is present only in the $\vel {\cal M}_2 \ver^2$ contribution 
is due to the soft photon emission from charged lepton line.
On the other hand, $\vel {\cal M}_1 \ver^2$ and 
$\mbox{\rm Re}\left[ {\cal M}_1 {\cal M}_2^\dagger \right]$ terms are free of 
this singularity. 
In this limit the $B_c \rar \tau \bar \nu_\tau \gamma$ decay can not
distinguished from the $B_c \rar \tau \bar \nu_\tau$ decay. In order to
obtain a finite result for the decay width, we must consider both decays
together. The infrared singularity arising from the  $\vel {\cal M}_2
\ver^2$ contribution must be canceled with $O(\alpha)$ virtual correction to the 
$B_c \rar \tau \bar \nu_\tau$ decay. In this work, our consideration is
slightly different, namely, the
$B_c \rar \tau \bar \nu_\tau \gamma$ process is not considered 
as a $O(\alpha)$ correction to the $B_c \rar \tau \bar \nu_\tau$ decay, but
rather a separate decay channel with hard photon radiation. Therefore we
impose a cut value on the photon energy, which will set an
experimental limit on the minimum detectable photon energy. We will consider
the case for which the photon energy threshold is larger than $50$, MeV
i.e., $E_\gamma \ge a \,m_{B_c}$, where $a \ge 0.01$. Using Eqs. (14)--(19)
and imposing the cut as the lower bound of the photon energy, the total
decay width takes the following form:
\bea
\Gamma &=& \frac{G^2 \alpha m_{B_c}^3}{64 \pi^2} \vel V_{cb} \ver^2 
\Bigg\{ \frac{1}{3} \int_\delta^{1-r} dx  
\left[ \vel f_1(x) \ver^2 + \vel f_2(x) \ver^2 \right] 
\frac{1}{(1-x)^2} \, x^3 (2 + r - 2 x) (1-r-x)^2 \nnb \\
&+& \!\! 4 f_{B_c}^2 \int_\delta^{1-r} dx \,\frac{r}{x(1-x)} 
\Bigg[ -4 + 8 r - 4 r^2 + 10 x - 14 r x + 4 r^2 x - 9 x^2 + 7 r x^2
+ 3 x^3 \nnb \\
&+& \!\! (1-x) (2 -2 r^2 - 3 x + rx + 2 x^2 ) 
\ell n \ga \frac{1-x}{r} \dr \Bigg]\nnb \\
&-& \!\! 4 f_{B_c}\int_\delta^{1-r} dx \,\frac{r x}{1-x} \Bigg[
(1-r-x) \Big( f_1(x) x + f_2(x) (1+r-2 x) \Big) \nnb \\
&-& \!\! (1-x) \Big( f_1(x) x + f_2(x) (2-r) \Big) 
\ell n \ga \frac{1-x}{r} \dr \Bigg] 
\Bigg\}~,
\eea           
where $x = 2 E_\gamma/m_{B_c}$is the dimensionless photon energy, 
$r = m_\tau^2/m_{B_c}^2$ and $\delta = 2 a$.

\section{NUMERICAL ANALYSIS}
It follows from Eq. (20) that, to be able to calculate the decay width,
explicit forms of the form factors $f_1$ and $f_2$ are needed. These form
factors are calculated in the framework of the light cone QCD sum rules
in \cite{R11}, and it was shown that, to a very good accuracy, 
their $p^2$ dependence can be represented in following pole forms:
\bea
f_1(p^2) = \frac{f_1(0)}{1 - p^2/m_1^2}~,~~~~~~~~~~ 
f_2(p^2) = \frac{f_2(0)}{1 - p^2/m_2^2}~,
\eea
where 
\bea
f_1(0) = 0.44 \pm 0.04~ \mbox{\rm GeV}~, ~~~~~ m_1^2 = 43.1 ~\mbox{\rm GeV}^2~,\nnb \\
f_2(0) = 0.21 \pm 0.02~ \mbox{\rm GeV}~, ~~~~~ m_2^2 = 48.0 ~\mbox{\rm GeV}^2~.\nnb
\eea
The following set of parameters have been used in the numerical analysis:
$f_{B_c} = 0.35$ GeV \cite{R16}--\cite{R18}, 
$V_{cb} = 0.04$ \cite{R19}, $\tau (B_c) = 0.46 \times 10^{-12}~s$ \cite{R1}. 
In Figs. (1) and (2), we present the photon energy dependence of the
branching ratio for two different fixed cut values, i.e., $\delta = 0.016$ and
$\delta = 0.032$. The steep increase in the value of the branching ratio at
small photon energies, is due to the Bremstrahlung part. 
Using these values, we have calculated the 
branching ratios and presented the results in Table 1.



\def\bos{\lower 0.2cm\hbox{{\vrule width 0pt height 0.65cm}}}
\def\aaa{\lower 0. cm\hbox{{\vrule width 0pt height .55cm}}}
\def\dol{\lower 0.4cm\hbox{{\vrule width 0pt height .6cm}}}


\begin{table}[h]
\begin{center}
\begin{tabular}{|c|c|c|}
\hline
                    &\bos $ {\bf \delta} = 0.016 $ & ${\bf \delta} = 0.032$ 
\\  \hline  \hline
{\bf Structure}     & \aaa $7.24 \times 10^{-6}$ & $7.24 \times 10^{-6}$ \\
\dol{\bf  dependent part} &          & \\
\hline      
{\bf Bremstrahlung} & \aaa $8.60 \times 10^{-5}$ & $6.53 \times 10^{-5}$ \\
\dol {\bf part}     &          &\\
\hline      
{\bf Interference}  & \aaa $2.17 \times 10^{-6}$ & $2.18\times 10^{-6}$ \\
\dol {\bf part}     &          & \\
\hline      
{\bf Total}         & \bos $9.54 \times 10^{-5}$ & $7.47 \times 10^{-5}$ \\
\hline      
\end{tabular}
\vskip 0.3 cm
\caption{}  
\end{center}
\end{table} 
            

From this table we observe that the main contribution to the $B_c \rar \tau
\bar \nu_\tau \gamma$ decay comes from the Bremstrahlung part and contributions
arising from the structure dependent and interference terms are comparable
to each other.

Note that the QCD sum rules prediction of the branching ratio of the 
$B_c \rar \ell \bar \nu_\ell \gamma~(\ell = e,~\mu)$ decay is \cite{R11} 
\bea
{\cal B}( B_c \rar \ell \bar \nu_\ell \gamma) \simeq 1.0 \times 10^{-5}~.
\eea

For completeness we present below the predictions of the branching ratios for
pure leptonic decays \cite{R15}:
\bea
{\cal B} \ga B_c \rar e \bar \nu_e \dr &=& 1.25 \times 10^{-9} \nnb \\
{\cal B} \ga B_c \rar \mu \bar \nu_\mu \dr &=& 5.26 \times 10^{-5} \nnb \\
{\cal B} \ga B_c \rar \tau \bar \nu_\tau \dr &=& 1.29 \times 10^{-2}~.
\eea
From a comparison of the results listed in Table 1 and Eq. (22), we observe
that the branching ratio of the radiative $\tau$ lepton decay is
higher than the corresponding ${\cal B}( B_c \rar \ell \bar \nu_\ell \gamma)$ and
the pure light leptonic decays.

Few words about the number of expected events at LHC are in order. As we
have noted earlier, approximately $\sim 2 \times 10^8~B_c$ mesons will 
be produced at LHC per year. Using the numerical result for the branching ratio of the 
${\cal B}( B_c \rar \tau \bar \nu_\tau \gamma)$ decay, for the number of
expected events at LHC, we get
\bea
N &\simeq& 2 \times 10^4~~(\mbox{\rm for}~~\delta = 0.016)~~\mbox{\rm and}~, \nnb \\
  &\simeq& 1.5 \times 10^4~~(\mbox{\rm for}~~\delta = 0.032)~. \nnb
\eea
Even with an efficiency of $\sim 10^{-2}$ in detecting the $\tau$ lepton,
approximately $~\sim 200$ events are expected to be observed. So, this decay
has a good chance to be detected at LHC.     

\newpage

\section*{FIGURE CAPTIONS}
FIG. 1. The dependence of the branching ratio of the 
$B_c \rar \tau \bar \nu_\tau \gamma$ decay on the dimensionless photon
energy. The lower bound of the dimensionless photon energy is taken as 
$\delta = 0.016$. \\ \\
FIG. 2. The same as FIG. 1, but at $\delta = 0.032$.

\begin{figure}
\vspace{25.0cm}
    \includegraphics{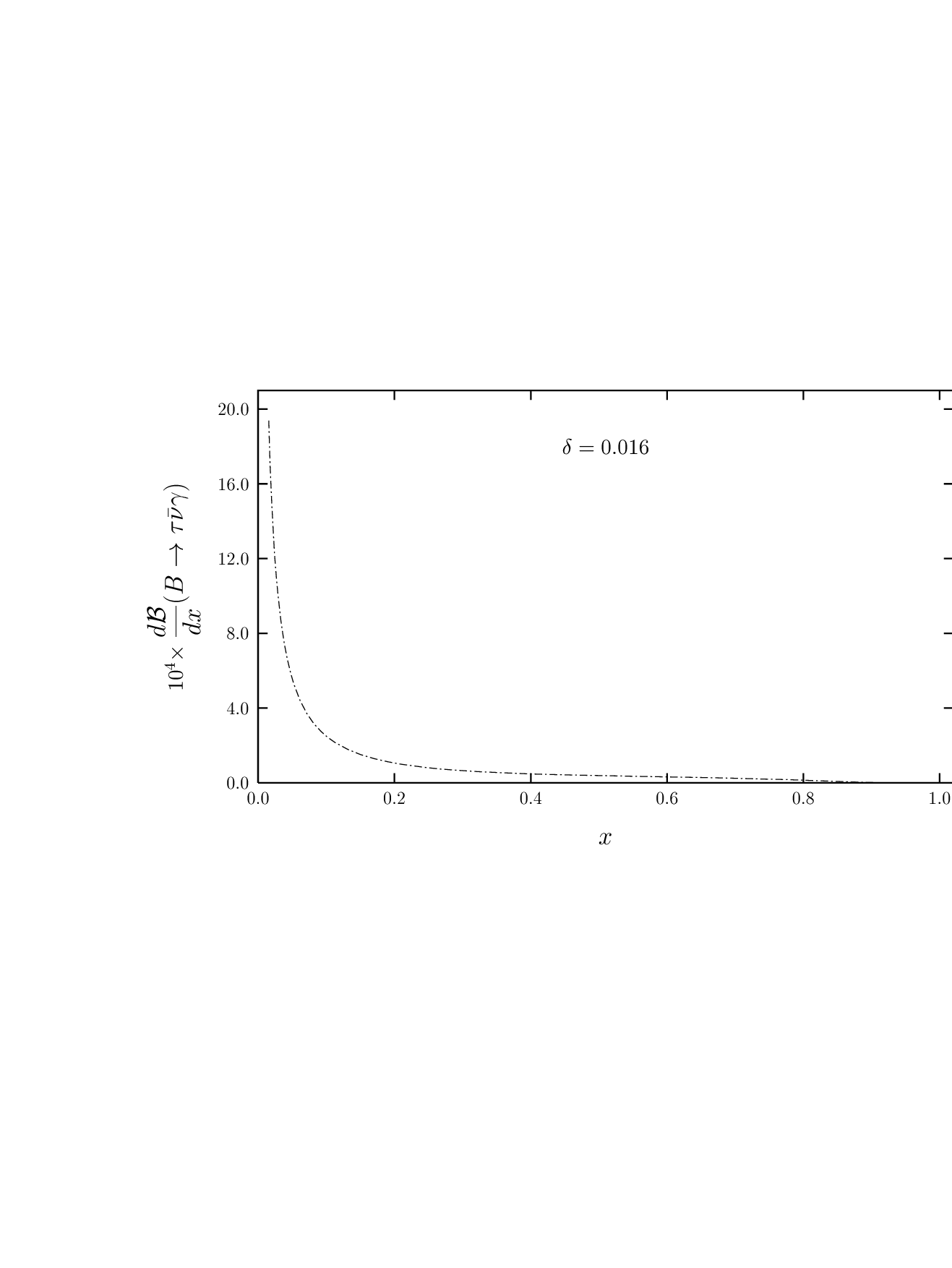}
\vspace{-17.cm}
\caption{ }
\vspace{16.5cm}
    \includegraphics{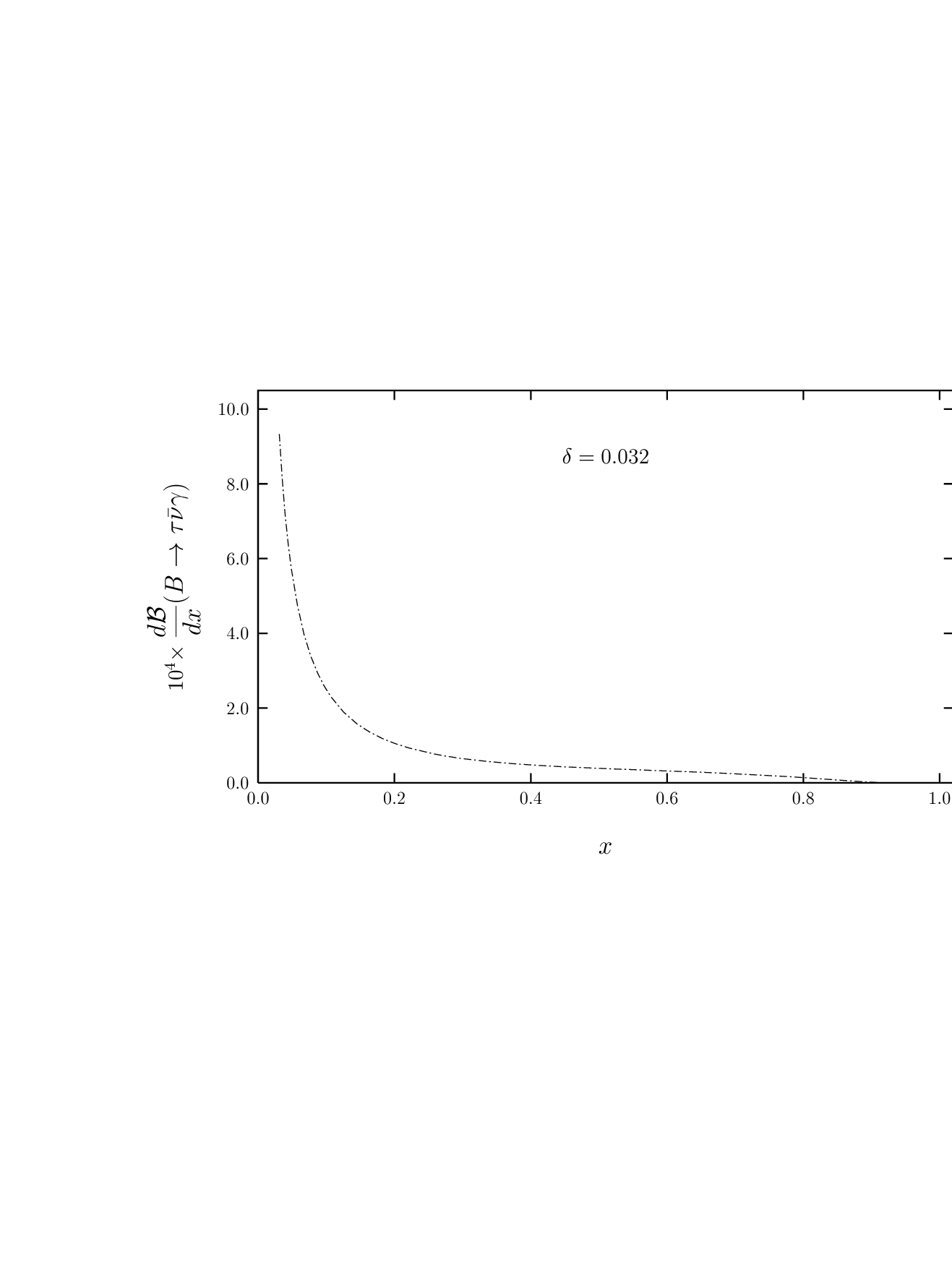}
    \vspace{.0cm}
\vspace{-5.cm}
\caption{ }

\end{figure}

\end{document}